\documentclass[reprint,superscriptaddress]{revtex4-2}

\usepackage{soul}
\usepackage{graphicx}
\usepackage{dcolumn}
\usepackage{bm}
\usepackage{xcolor}
\usepackage{hyperref}
\usepackage[mathlines]{lineno}
\usepackage{amsmath}

\begin{document}

\preprint{APS/123-QED}

\title{Pushing-Induced Arrest Across Lattices and Dimensions}

\author{I. Shitrit }
\email{itamarshtrit@mail.tau.ac.il}

\affiliation{%
School of Chemistry, Center for the Physics and Chemistry of Living Systems, 
and the Sackler Center for Computational Molecular Materials Science, Tel Aviv University, 6997801 Tel Aviv, Israel
}%
\author{O. Lauber Bonomo}
\email{o.lauber@nyu.edu
}
 \affiliation{%
School of Chemistry, Center for the Physics and Chemistry of Living Systems, 
and the Sackler Center for Computational Molecular Materials Science, Tel Aviv University, 6997801 Tel Aviv, Israel
}%
\affiliation{
 Center for Urban Science and Progress, New York University Tandon
School of Engineering, Brooklyn 11201, New York, United States of
America.
}%
\author{S. Reuveni }%
\email{shlomire@tauex.tau.ac.il
}
\affiliation{%
School of Chemistry, Center for the Physics and Chemistry of Living Systems, 
and the Sackler Center for Computational Molecular Materials Science, Tel Aviv University, 6997801 Tel Aviv, Israel
}%

\date{\today}

\begin{abstract}

Tracer–media interactions can give rise to transport phenomena beyond classical models; e.g., obstacle pushing can eliminate percolation. We demonstrate that the existing ``snowplow'' mechanism proposed to explain this effect fails in 3D. We show that confinement is governed by emergent trapping—rare ``door-closing'' events that occur with an approximately constant probability per step at low obstacle densities, thus yielding exponential survival. This allows prediction of the time-dependent mean-squared displacement from short-time estimates of the diffusion constant and trapping probability, providing a minimal description of pushing-induced arrest across lattices and dimensions.
\end{abstract}

\maketitle

\begin{figure}[t]
\includegraphics{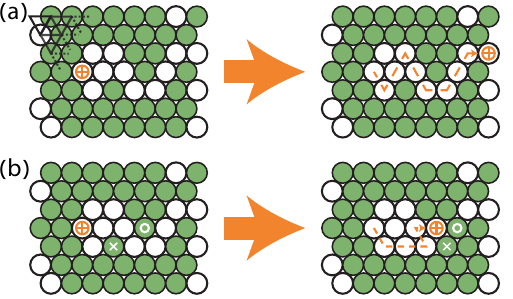}

\caption{\label{fig: walk example} \textbf{AIL vs. \textit{Sokoban} random walk.} Examples of an (a) AIL and (b) \textit{Sokoban} random walk on the Triangular lattice. Green nodes are occupied with obstacles, and white nodes are free. The orange tracers wander around obeying their respective laws of motion. While the AIL escapes, the \textit{Sokoban} tracer happens to permanently confine itself. Two obstacles are marked with a circle and a cross, for traceability. They are moved by the \textit{Sokoban} tracer during the walk, but are identical to all other obstacles.}
\end{figure}

\noindent From advancing a chess pawn toward promotion to navigating busy crowds, pushing is synonymous with forward motion. Despite the ubiquity of this heuristic in daily life, classical models describing the motion of tracers through disordered environments typically overlook pushing dynamics. A canonical example of this ruling paradigm is the “ant in a labyrinth” (AIL) model \cite{ de1976percolation, stauffer2018introduction, ben2000diffusion}. There, an ‘ant’ is depicted as a random walker, surrounded by immovable obstacles as illustrated in Fig. \ref{fig: walk example}(a). This disparity between the static assumptions of traditional percolation models \cite{shante1971introduction, essam1980percolation, saberi2015recent, li2021percolation} and the pushy nature of motion in the real-world raises fundamental questions: How does pushing affect transport in disordered media? And does it always translate to forward motion?

Recently, these questions have been studied in controlled experiments. Biswas et al., examined the stochastic transport of a self-propelled camphor boat navigating a field of passive paper discs floating on water \cite{biswas2020first}. The mean first passage time, defined as the average travel time from the center to the boundary of the arena, increased monotonically with the fraction of area occupied by the discs. In contrast, fluctuations in travel time displayed non-monotonic behavior, reaching lower values at intermediate area fractions.

Although obstacles hinder transport over short timescales, the retention of memory from previous trajectories can facilitate transport in the long term. Altshuler et al., investigated the movement of a self-propelled bristle robot within a field of movable obstacles. By directly modifying the positions of obstacles in its path, the tracer creates trails \cite{Altshuler2024}. These trails give rise to a form of environmental memory that expedites search by increasing the effective diffusion coefficient. Similar environmental memory was observed by Dias et al. on a microscopic scale \cite{dias2023environmental}. By employing light-activated Janus particles mixed with passive silica colloids, they demonstrated that environmental memory can arise in a system of “clueless” individuals, promoting group formation. Collectively, these findings indicate that the dynamics of active tracers are fundamentally different when obstacles can be pushed.

Motivated by these findings, Bonomo \& Reuveni introduced the \textit{Sokoban} random walk, a minimal model aimed at capturing the effect of obstacle pushing \cite{Lauber2023}. In the model, a random walker on a lattice is allowed to push at most one obstacle at a time; see Fig. \ref{fig: walk example}(b). While the ability to push obstacles was expected to allow the \textit{Sokoban} to venture further away, the model was used to demonstrate that this is not always true. In fact, it was found that the \textit{Sokoban} always ends up confined. The region visited prior to confinement has a characteristic size, which is set by the initial obstacle density. These features contrast with the AIL model, where a sharp transition between restricted motion and free diffusion occurs at a specific critical density $0 < \rho_{c} < 1$ \cite{stauffer2018introduction}.

\par The surprising loss of the percolation transition due to obstacle pushing raises the question of whether this effect is geometry-specific or persists more generally. This question was subsequently addressed, revealing that on the infinite-dimensional Bethe lattice, the \textit{Sokoban} undergoes a percolation transition, occurring at a critical density higher than its non-pushy counterpart, as intuition dictates \cite{Lauber2024}. This critical density was also found by Vandewalle et al. who considered a fluid version of the model in the context of epidemics \cite{VANDEWALLE19961}. 

Tracer–media interactions can thus qualitatively and quantitatively reshape transport in a topology- and dimension-dependent manner. This calls for a unifying description that remains valid across geometries. Here, we investigate the \textit{Sokoban} model across lattices and dimensions and show that confinement is governed by emergent trapping—rare, irreversible “door-closing” events—which yields a minimal coarse-grained description of pushing-induced arrest.

\par \textit{The model---}We consider the \textit{Sokoban} model, in which a tracer with limited pushing ability performs a random walk that, among other things, reconfigures its environment. The medium is represented as a lattice where each node is randomly occupied by an obstacle with probability $\rho$. A single tracer is initialized on the lattice and proceeds via a sequence of feasible moves. A feasible move for the tracer satisfies one of two conditions: (i) stepping into a vacant neighboring node; or (ii) pushing a single obstacle in its direction of motion, provided that the next site is unoccupied. Pushing more than a single obstacle at a time is forbidden. At each time step, the tracer selects a move at random from the current set of feasible moves. For an illustration of such a walk, see Fig. \ref{fig: walk example} (b). 

\par Previous work explored the \textit{Sokoban} model on the 2D Square lattice \cite{Lauber2023}. It was found that the mean squared displacement (MSD) of the walk saturates to a finite value (see appendix \ref{appendix: snowplow 2D})
\begin{equation} \label{eq: snowplow prediction} 
\mathrm{MSD}_\infty=\left(\frac{1-\rho}{C\rho} \right)^{\frac{2}{\gamma}},
\end{equation}
after many steps. Equation (\ref{eq: snowplow prediction}) relates $\mathrm{MSD}_\infty$, the saturation $\mathrm{MSD}$ value, and the obstacle density $\rho$ via the positive constants $C$ and $\gamma$. These constants are set by the fractal nature of the walk. Specifically, the authors defined: (i) $\mathcal{A}$, the number of distinct nodes the tracer visits during the walk; and (ii) $\mathcal{P}$, the number of unvisited nodes, that are within two steps of a visited node, reflecting the region of interest when considering feasible moves. It was shown that the averages of these quantities obey a fractal scaling 
\begin{equation} \label{eq: snowplow definitions} 
\langle\mathcal{A}\rangle \simeq A\sqrt{\mathrm{MSD}_\infty^\alpha}, \quad
\langle\mathcal{P}\rangle \simeq B\sqrt{\mathrm{MSD}_\infty^\beta}.
\end{equation}
Following these definitions, the constants in Eq. (\ref{eq: snowplow prediction}) are given by $\space \gamma=\alpha-\beta$ and $C=f \frac{A}{B}$, where $f$ is a factor relating to occupied nodes in $\mathcal{A}$ and unoccupied nodes in $\mathcal{P}$ at the end of the walk (see appendix \ref{appendix: snowplow 2D}).

\par \textit{The ``snowplow effect"---}To elucidate the physical mechanism behind the emergent confinement in the 2D Square lattice, picture plowing snow when clearing your yard. As the snow is pushed aside, it accumulates at the perimeter of the cleared area. Since the cleared area grows faster than its perimeter, the displaced snow will pile up at the rims. Ultimately, the periphery becomes filled with heavy snow piles that are difficult to move. The \textit{Sokoban} walk displays a similar behavior, which we henceforth refer to as the ``snowplow effect". As the tracer ventures around, it clears obstacles from visited nodes. These obstacles do not vanish, but are rather pushed together into the perimeter of the walk. For the \textit{Sokoban} walk, $\gamma$ was found to be positive \cite{Lauber2023}, indicating that the area grows faster than the perimeter. Therefore pushed obstacles inevitably form an impassable, double-layered, wall which confines the walk. 

\par On the 2D Square lattice Eq. (\ref{eq: snowplow prediction}), which was formulated based on the ``snowplow effect", successfully predicts the empirical $\mathrm{MSD}_\infty$ \cite{Lauber2023}. However, it is unclear whether this is true for other lattices. In fact, little is known about the \textit{Sokoban} random walk outside the 2D Square lattice. Does it always confine itself? And, if so, is the ``snowplow effect" the mechanism behind confinement? We now address these questions by studying the \textit{Sokoban} on the Triangular and Hexagonal (honeycomb) lattices \cite{stauffer2018introduction}, testing the predictions of Eq. (\ref{eq: snowplow prediction}). Throughout this work, we omit high obstacle densities, and instead focus on densities where the tracer is able to explore far beyond its immediate surroundings before confinement occurs, i.e., $\sqrt{\mathrm{MSD}_\infty}\gg1$, allowing us to probe emergent large-scale phenomena rather than trivial local exploration.

\par
\begin{figure}[t]
\includegraphics[width=0.48\textwidth]{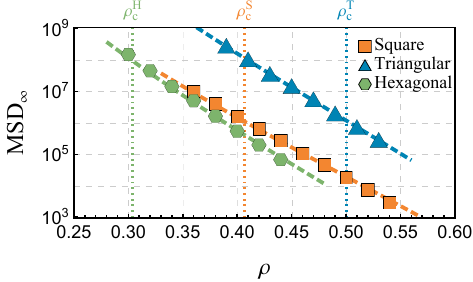}
\caption{\textbf{Existence of ``Snowplow effect" for the \textit{Sokoban} in different 2D lattices.} The terminal MSD, $\mathrm{MSD}_{\infty}$, vs. the obstacle density $\rho$ for the Square, Triangular and Hexagonal lattices. Predictions coming from Eq. (\ref{eq: snowplow prediction}) (dashed lines) are in excellent agreement with Monte Carlo simulations (markers). The dotted vertical lines represent the percolation thresholds for the AIL (no obstacle pushing) in the respective lattices.}
\label{fig: 2D fractal prediction}
\end{figure}

 In Fig. \ref{fig: 2D fractal prediction}, we compare $\mathrm{MSD}_\infty$ values coming from Monte-Carlo simulations (symbols) to predictions coming from Eq. (\ref{eq: snowplow prediction}) (dashed lines). To obtain the latter, we first estimated $C$ and $\gamma$ via Eq. (\ref{eq: snowplow definitions}) (see appendix \ref{appendix: snowplow 2D} for simulation details and results). We performed simulations for the three lattices, with obstacle densities both above and below their respective percolation thresholds.

Below the percolation threshold, the $\mathrm{MSD}$ of the AIL diverges as the number of steps grows. Conversely, for the \textit{Sokoban} we find a finite $\mathrm{MSD}_\infty$. Furthermore, we see that the predictions of Eq. (\ref{eq: snowplow prediction}) are in excellent agreement with simulations. 
At very low obstacle densities computing limitations prevent simulations. There, Eq. (\ref{eq: snowplow prediction}) predicts that for $\gamma>0$ and $\rho>0$ a finite $\mathrm{MSD}_\infty$ is obtained. This suggests that pushing eliminates percolation completely at any obstacle density, by means of the ``snowplow effect". 

\par Surprisingly, confinement by the ``snowplow effect" does not extend to all lattices. For example, in the Bethe lattice pushing facilitates percolation \cite{Lauber2024} and the ``snowplow effect" is notably absent. As the Bethe lattice is effectively infinite-dimensional \cite{stauffer2018introduction}, the absence of a ``snowplow effect" there hints that its emergence is dimension dependent. To probe into this matter, we cranked up the dimension knob, and simulated the \textit{Sokoban} random walk on the 3D Simple Cubic (SC) lattice. We measured the fractal dimensions of the volume and surface area which are the 3D equivalents of the $\mathcal{A}$ and $\mathcal{P}$ used in Eq. (\ref{eq: snowplow definitions}). Finally, we used these to generate a prediction of the $\mathrm{MSD}_\infty$ based on Eq. (\ref{eq: snowplow prediction}). This procedure, that provided accurate estimations in 2D, fails for the 3D SC lattice. While we still observe condensation of obstacles along the perimeter of the walk (see appendix \ref{appendix: condensation}), the tracer is confined well before reaching the predicted $\mathrm{MSD}_\infty$. In fact, the predicted $\mathrm{MSD}_\infty$ is orders of magnitude larger than the one obtained from simulations (see appendix \ref{appendix: 3d fractal}).

\begin{figure*}[t]
\includegraphics[width=0.96\textwidth]{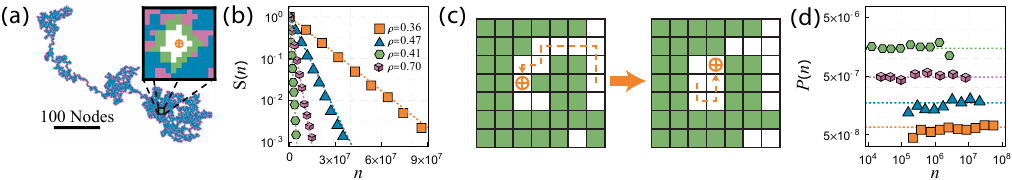}
\centering
\caption{\textbf{Trapping of the \textit{Sokoban} random walk.} (a) The \textit{Sokoban} random walk ends its life in a pocket (zoom-in) that can be orders of magnitude smaller than the size of the walk. Here, the area, $\mathcal{A}$, and perimeter, $\mathcal{P}$, that were defined below Eq. (\ref{eq: snowplow prediction}) are colored blue and pink, respectively. Obstacles confining the terminal pocket are colored green. (b) The survival probability, $S(n)$, of a \textit{Sokoban} tracer for various lattices. Dashed lines are exponential fits. Additional densities are presented in appendix \ref{appendix: trapping data}. (c) An example of confinement via trapping. On the left, the tracer enters a pocket of obstacles, and then ``closes the door", permanently confining itself to the six nodes inside the pocket. (d) Comparison of the directly measured trapping probability (markers) and the exponential decay rate (dashed lines) extracted from the fits in panel (b).}
\label{fig: trapping explained}
\end{figure*}

\par This mismatch indicates that in 3D, confinement is not controlled by the eventual macroscopic buildup of a snowplowed wall. Rather, there appears to be a competing mechanism that arrests the walk earlier. To characterize the kinetics of this arrest, we must look beyond $\mathrm{MSD}_\infty$, which gives a characteristic size for the walk. To do so, we shift our attention to the time it takes the walk to become confined.

\par \textit{Trapping approach---}When examining the behavior of the \textit{Sokoban} random walk over time, two distinct phases are identified. Initially, the walk explores the lattice, randomly pushing obstacles. At some point, the walk stops discovering new nodes. When this happens, it is confined to a permanent set of nodes. Surprisingly, this terminal cage is orders of magnitude smaller than the total area visited by the walk, see Fig. \ref{fig: trapping explained}(a). This has also been observed in \cite{Lauber2023} and \cite{Singh2026}. 
\par Next, we examine this unexplained phenomenon. We define the survival probability, $S(n)$, that the walk has not entered its terminal cage for the last time by step $n$. We estimate this probability using simulations on the Square, Triangular, Hexagonal and Cubic lattices (simulation details in appendix \ref{appendix: trapping data}). Results in Fig. \ref{fig: trapping explained}(b) and appendix \ref{appendix: trapping data} show that the survival decays exponentially (to a good approximation) with the number of steps, $n$, across lattices and obstacle densities. This exponential decay can be explained using a trapping approach. 

\par First, we elucidate the trapping event. How does the walk transition from free exploration to the terminal state? Consider the left panel of Fig. \ref{fig: trapping explained}(c). The path depicted there is completely reversible. This means that the tracer can enter and leave the terminal cage multiple times before confinement occurs. However, shortly after the tracer enters the terminal pocket for the \textit{last} time an irreversible obstacle pushing event occurs. The tracer enters the pocket and ``closes the door" behind itself, see right panel of Fig. \ref{fig: trapping explained}(c). In practice, we set the trapping step as the last time the tracer entered the terminal pocket.

\par Now that we have established how trapping occurs, we can try and explain the behavior of $S(n)$. We denote the probability of the tracer to confine itself at step $n$ as $P_n$. The survival is then given by $S(n)=\prod_{i=1}^n1-P_i$. In Fig. \ref{fig: trapping explained}(d), we estimate $P_n$ from simulations for the bulk of the survival distribution (see appendix \ref{appendix: trapping data} for additional details). We find that it is approximately constant, i.e., does not depend on $n$, and very small. We thus approximate $S(n)\propto e^{-Pn}$. The decay rates measured in Fig. \ref{fig: trapping explained}(b) are also displayed as dashed lines, showing good agreement with direct estimates of $P$ (symbols). 

\par The trapping approach not only captures the survival kinetics, it can also be used to predict the evolution of $\mathrm{MSD}(n)$. To derive an equation for this quantity, we write $\mathrm{MSD}(n)$ as a sum of two contributions. The first comes from tracers that have not yet been confined by step $n$. We observe that for such tracers $\mathrm{MSD}(n|\text{survived})=2Dn$, with $D$ being $\rho$ dependent (for additional details see appendix \ref{appendix: trapping derivation}). Thus, their contribution to the total $\mathrm{MSD}(n)$ is $ 2Dn e^{-nP}$. The second contribution comes from tracers that were confined prior to step $n$. We again use the law of total probability to integrate over the contributions of tracers that were confined at each $n'<n$. This integral yields $2D\left(\frac{1}{P} -\left(\frac{1}{P}+n \right)e^{-nP} \right)$. Summing the two contributions we obtain (see appendix \ref{appendix: trapping derivation})
\begin{align}
\label{eq: MSD _t_ analytical}
\mathrm{MSD}(n) \approx2D(1-e^{-nP})/P.
\end{align} 

\begin{figure}[t]
\includegraphics[width=0.48\textwidth]{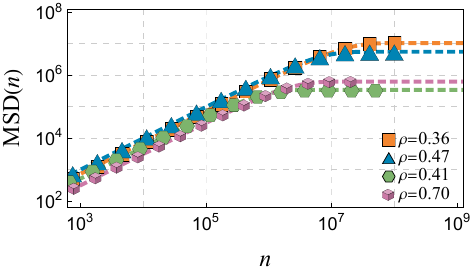}
\caption{\textbf{Successful prediction of $\mathrm{MSD}(n)$ across lattices and dimensions.} The MSD of the \textit{Sokoban} random walk (markers for various lattices) compared to the prediction provided by Eq. (\ref{eq: MSD _t_ analytical}) shown in dashed lines. Additional densities presented in appendix \ref{appendix: trapping data}.}
\label{fig: MSD _t_}
\end{figure}

Equation (\ref{eq: MSD _t_ analytical}) enables prediction of the macroscopic ensemble property $\mathrm{MSD}(n)$ using two microscopic parameters: the trapping probability, $P$, and diffusion constant $D$. Crucially, these parameters can be easily estimated from the short-time dynamics to predict $\mathrm{MSD}(n)$ at all times, which was previously impossible. Doing so for all four lattices, and using Eq. (\ref{eq: MSD _t_ analytical}) provides high accuracy prediction of $\mathrm{MSD}(n)$, see Fig. \ref{fig: MSD _t_} and appendix \ref{appendix: trapping data}. Thus, trapping succeeds in capturing behavior in both 2D and in 3D, where the ``snowplow effect" failed. 

\par Trapping and the ``snowplow effect" are two competing mechanisms of confinement. While the former, describes a long and continual process, the latter is an instantaneous one. Thus, it is possible that a tracer will trap itself while ``snowplowing". In 2D, trapping typically does not occur until the macroscopic snowplowed cage is essentially established; as a result, the two descriptions agree on $\mathrm{MSD}_\infty$, explicitly linking the macroscopic parameters $(C,\gamma)$ that control the snowplow prediction in Eq. (\ref{eq: snowplow prediction}) to the microscopic kinetic parameters $(D,P)$ that control the trapping dynamics. Specifically, taking $n\to\infty$ in Eq. (\ref{eq: MSD _t_ analytical}) and comparing with Eq. (\ref{eq: snowplow prediction}) we have: $\frac{2D}{P}=\left(\frac{1-\rho}{C\rho} \right)^{\frac{2}{\gamma}}$. In 3D, trapping arrests the walk before a macroscopic wall can form, so the snowplow estimate becomes a geometric upper bound rather than the realized terminal $\mathrm{MSD}$.

\par \textit{Classical vs. \textit{Sokoban} trapping. Exponential kinetics revisited---}\textit{Sokoban} trapping is distinct from the classical trapping problem \cite{klafter2011first,Redner2010}. There, a tracer wanders around in an arena which is randomly populated with traps that annihilate the tracer upon contact. Since annihilation can only occur when a newly discovered site turns out to be a trap, the number of distinct sites visited is key to understanding kinetics in this problem. Conversely, in the \textit{Sokoban}, trapping is an emergent property, and obstacle configurations become traps only when obstacle pushing is introduced. Moreover, the tracer can get trapped even when stepping into a site that has already been discovered in the past (see discussion on the example in Fig. \ref{fig: trapping explained}(c)). In addition, unlike classical trapping, a \textit{Sokoban} tracer can destroy traps, e.g., by pushing the free obstacle in Fig. \ref{fig: trapping explained}(c) one site downwards, thus neutralizing the trap indefinitely. We therefore argue that pushing promotes a new, self-generated, trapping mechanism that goes beyond the existing theory of classical trapping.

For classical trapping, at short--intermediate time scales, one finds exponential survival kinetics, $S(n)\propto e^{-an}$, in 3D and exponential kinetics with logarithmic corrections, $S(n)\propto e^{-an/ln(n)}$, in 2D \cite{klafter2011first}. At asymptotically long times, survival is governed by rare initial configurations of the traps, giving rise to stretched exponential survival $S(n)\propto e^{-an^{d/d+2}}$ \cite{donsker1979number,klafter2011first,Redner2010}, where $d$ is the dimension. Very recently, Singh et al. proved that the long-time asymptotics of the survival of the \textit{Sokoban} on the 1D lattice is $S(n)\propto \mathcal{K}(n)e^{-an^{1/3}}$ with $\mathcal{K}(n)\propto n^{7/6}$ \cite{Singh2026,singh2026-2}. This modified stretched exponential asymptotics, reminiscent of classical trapping, led Singh et al. to postulate $S(n)\propto \mathcal{K}(n)e^{-an^{1/2}}$ asymptotics for the \textit{Sokoban} on the 2D square lattice.

Singh et al. focused on obstacle densities higher than those considered here when characterizing the survival kinetics. For the 2D square lattice, we focused on $\rho\leq0.4$, while \cite{Singh2026} presents results for $\rho=0.5, 0.6$. Performing detailed simulations of the \textit{Sokoban} for these higher densities ($10^6$ walk realizations vs. $10^4$ above), we find full agreement with the results in \cite{Singh2026}, affirming strong deviations from exponential kinetics (see appendix \ref{appendix: comparison}). However, as the obstacle density is lowered, deviations from exponential decay become progressively less pronounced, such that for the densities considered here, exponential decay provides an almost perfect description of the survival kinetics: emerging after a short transient and persisting out to times nearly nine standard deviations beyond the mean ($S(n)\approx10^{-5}$, see appendix \ref{appendix: low-density}). These findings call for the development of a theory that will be able to explain \textit{Sokoban} survival kinetics in dimensions two and up, possibly revealing a density dependent crossover between exponential kinetics and the modified stretched-exponential asymptotics postulated by Singh et al.

\par \textit{Conclusions and outlook---}We studied the \textit{Sokoban} random walk, a minimal model for transport in the presence of tracer-media interactions, across lattices and dimensions. In two-dimensional Euclidean tilings, we find that the exploration length scale is well described by Eq. (\ref{eq: snowplow prediction}) and the macroscopic ``snowplow'' picture: obstacles cleared from the explored region accumulate at its perimeter and ultimately form an impenetrable wall, yielding a finite $\mathrm{MSD}_\infty$ for any $\rho>0$. In three dimensions, this macroscopic reasoning breaks down: despite boundary condensation, confinement occurs at length scales that can be orders of magnitude smaller than the snowplow estimate, indicating an additional arrest mechanism.

At high obstacle densities, the walker is either completely confined from the get-go, or has very limited ability to explore beyond its immediate surroundings. We thus focused on lower obstacle densities for which $\sqrt{ \mathrm{MSD}_{\infty}}\gg1$. There, we identified the missing mechanism as \emph{emergent trapping}. Confinement is triggered by rare, irreversible local rearrangements that close the ``door'' behind the tracer. At low obstacle densities, the resulting trapping probability is approximately constant in step number, leading to exponential survival. This perspective yields a closed-form prediction for the time dependent Mean-Squared-Displacement. The prediction is encapsulated in Eq. (\ref{eq: MSD _t_ analytical}), in terms of two microscopic quantities accessible from short-time dynamics: the diffusion coefficient $D(\rho)$ and the instantaneous trapping probability $P(\rho)$. Trapping and the “snowplow” effect are complementary mechanisms that together provide a coherent picture of confinement. In this unified view, snowplowing provides a geometric upper bound on the ultimate size of the explored region, whereas trapping provides the kinetic route to arrest. This distinction becomes decisive in 3D, where transience prevents the sustained boundary processing required for snowplow-controlled confinement.

A key implication is that an interacting, history-dependent process can nevertheless admit a sharp coarse-grained description. Indeed, recent experiments and models of environmental memory captured the effect of pushing via an effective diffusion coefficient $D(\rho)$ \cite{Altshuler2024}. Our results show that this is generally insufficient: pushing also generates a distinct kinetic pathway to arrest through emergent trapping. Once the pair $\{D(\rho),P(\rho)\}$ is known from short-time dynamics, Eq.~(\ref{eq: MSD _t_ analytical}) determines the subsequent mean-squared displacement, providing a minimal predictive characterization of transport despite the underlying complexity.

More broadly, the \textit{Sokoban} walk serves as a counterexample in which pushing hinders, rather than enhances, transport by creating self-generated traps. This underscores that ``more pushing'' need not imply faster spreading, and that the outcome depends on the microscopic laws of motion. An important next step is therefore to predict $D(\rho)$ and $P(\rho)$, for pushing and other tracer–media interactions—from local geometry and rearrangement statistics, enabling systematic exploration of alternative laws of motion and delineating regimes where interactions eliminate percolation, restore it, or generate qualitatively new dynamics.

\textit{Acknowledgments---}This project has received funding from the European Research Council (ERC) under the European Union’s Horizon 2020 research and innovation program (grant agreement No. 947731).

\bibliography{bibl}%
\clearpage
\onecolumngrid
\appendix

\makeatletter
\@addtoreset{figure}{section}
\@addtoreset{equation}{section}
\makeatother

\renewcommand{\thefigure}{\thesection\arabic{figure}}

\section{Equation (1) and the snowplow effect}\label{appendix: snowplow 2D}

\begin{figure}[h]
\includegraphics[width=0.96\textwidth]{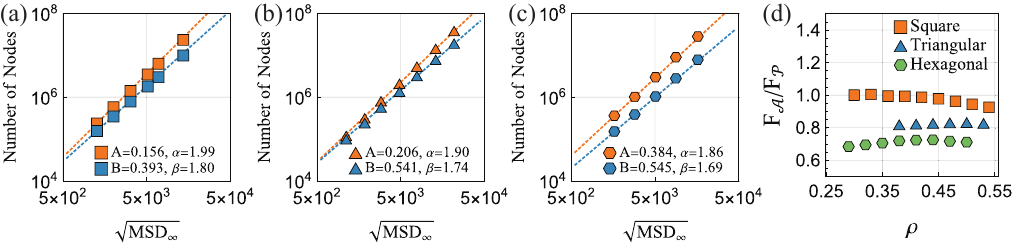}
\centering
\caption{\textbf{In 2D, $\mathcal{A}$ and $\mathcal{P}$ are fractals.} $\langle\mathcal{A}\rangle$ and $\langle\mathcal{P}\rangle$ from Eq. (\ref{eq: snowplow definitions}) for the \textit{Sokoban} on (a) Square (b) Triangular and (c) Hexagonal lattices. In (a-c), the orange dataset corresponds to the average area, $\langle\mathcal{A}\rangle$, while the blue corresponds to the average perimeter, $\langle\mathcal{P}\rangle$. We also present the measured ratio $F_\mathcal{A}/F_\mathcal{P}$ for these lattices in panel (d), where different markers correspond to different lattices.}\label{fig: 2D fractal appendix}
\end{figure}

\noindent We begin by providing a detailed explanations of the derivation performed in \cite{Lauber2023}. The authors derived Eq. (\ref{eq: snowplow prediction}) by starting with a condition that is based on the snowplow effect. They first noted that the following is a sufficient condition for caging:
\begin{equation} \label{eq: ofek snowplow}
\mathcal{A}\rho = \mathcal{P}(1-\rho).
\end{equation}
This condition requires that the obstacles in the area will be pushed to occupy all the free nodes in the perimeter. They then plugged Eq. (\ref{eq: snowplow definitions}) into Eq. (\ref{eq: ofek snowplow}), after taking averages there, to obtain Eq. (\ref{eq: snowplow prediction}). Yet, in practice, it was found that not all obstacles are evacuated from the area, and not all free nodes in the perimeter end up occupied. To account for that, Eq. (\ref{eq: ofek snowplow}) was modified to state that the \textit{Sokoban} gets caged when a fraction $F_\mathcal{A}$ of the obstacles that resided in visited nodes were pushed to occupy a fraction $F_\mathcal{P}$ of perimeter nodes that were initially vacant to get
\begin{equation} \label{eq: F corrections}
\mathcal{A}\rho F_\mathcal{A} = \mathcal{P}(1-\rho)F_\mathcal{P}.
\end{equation}
Interestingly, Eq. (\ref{eq: snowplow prediction}) remains valid, but with a redefined constant, $C$, that is now given by
$C= \frac{AF_\mathcal{A}}{BF_\mathcal{P}}$. 

We used Monte-Carlo simulations to estimate all the relevant parameters contributing to Eq. (\ref{eq: snowplow prediction}): $A,B,\alpha ,\beta, F_\mathcal{A}/F_\mathcal{P}$. We simulated \textit{Sokoban} random walks for a range of densities on each lattice (Square, Triangular, Hexagonal). For each density, we simulated $10^4$ walkers. Every point of data, is the ensemble average of each set of simulations.
The resulting values of all these simulations are presented in Fig. \ref{fig: 2D fractal appendix}. Specifically, in Fig. \ref{fig: 2D fractal appendix}(d), we show that on the Square lattice the ratio $F_\mathcal{A}/F_\mathcal{P}$ is very close to unity for all but the highest obstacle densities considered in this study. We thus take it to be unity. For both the Hexagonal and Triangular lattices, the ratio $F_\mathcal{A}/F_\mathcal{P}$ is roughly constant. Thus, when determining $C$ for these lattices, we use the average value obtained from simulations. The obtained constants, required for Eq. (\ref{eq: snowplow prediction}), will now be listed. Square: $\gamma^S=0.19, C^S=0.40$. Triangular: $\gamma^T=0.17, C^T=0.31$. Hexagonal: $\gamma^H= 0.17,C^H=0.50$.

\newpage

\section{Perimeter condensation on the Square and Simple Cubic lattices.}\label{appendix: condensation}

\begin{figure}[h]
\includegraphics[width=0.48\textwidth]{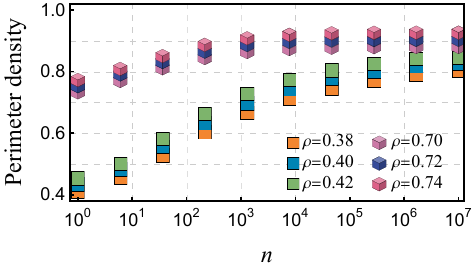}
\centering
\caption{ \textbf{Perimeter condensation on the Square and Simple Cubic lattices.} The density of obstacles on the perimeter of the \textit{Sokoban} random walk vs. the number of steps, for the 2D Square lattice (squares) and 3D Simple Cubic lattice (cubes).}
\label{fig: condensation}
\end{figure}

\noindent In Fig. \ref{fig: condensation}, we present the average density of obstacles in the perimeter throughout the walk for different densities and different lattices (Square and Simple Cubic, differentiated by corresponding symbols). In both lattices, we observe that as more steps are taken the density of obstacles on the perimeter increases.


\section{Breakdown of Eq. (1) for \textit{Sokoban} on the Simple Cubic lattice} \label{appendix: 3d fractal}

\begin{figure}[h]
\includegraphics[width=0.96\textwidth]{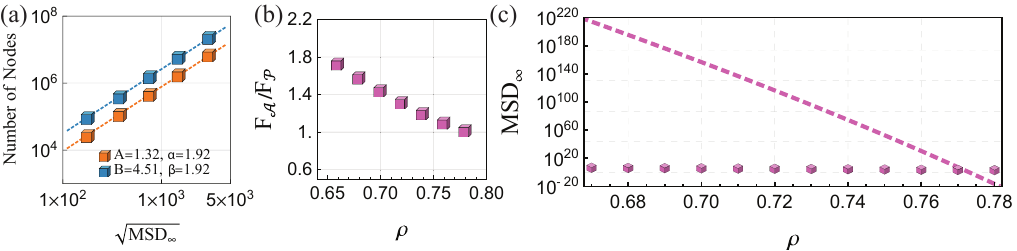}
\centering
\caption{\textbf{Breakdown of Eq. (1) for \textit{Sokoban} on the Simple Cubic lattice} (a) Measurements of the fractal dimensions on the Simple Cubic lattice. The orange dataset corresponds to the visited volume while the blue corresponds to the surface area. Note that we find $\gamma = \alpha-\beta\approx 2*10^{-3}$, which is very close to zero. (b) measurements of the ratio $F_\mathcal{A}/F_\mathcal{P}$. (c) The $\mathrm{MSD}_\infty$ for the \textit{Sokoban} random walk on the Simple Cubic lattice vs. the obstacle density. Predictions coming from Eq. (2) (dashed line) widely overestimate results coming from simulations (symbols).}
\label{Fig: 3D fractals}
\end{figure}

\noindent In the Simple Cubic lattice, we estimated the fractal dimensions of the volume and surface area which are the 3D equivalents of the $\mathcal{A}$ and $\mathcal{P}$ used in Eq. (\ref{eq: snowplow definitions}). We also estimated the ratio $F_\mathcal{A}/F_\mathcal{P}$. Results are shown in Fig. \ref{Fig: 3D fractals}(a) and Fig. \ref{Fig: 3D fractals}(b) respectively. The prediction generated by substituting these parameters into Eq. (1) is presented as a dashed line in Fig. \ref{Fig: 3D fractals}(c). It is clear that this prediction wildly overestimates simulation results for $\mathrm{MSD}_\infty$ on the Simple Cubic lattice (symbols).

\section{Exponential survival across lattices and obstacle densities}\label{appendix: trapping data}
In this appendix, we provide more data supporting our findings. We also explain how $P(n)$ was estimated directly. We show that \textit{Sokoban} trapping kinetics is exponential for additional densities in each lattice, and elaborate on the details of the simulation.

\par
\begin{figure}[t]
\includegraphics[width=0.96\textwidth]{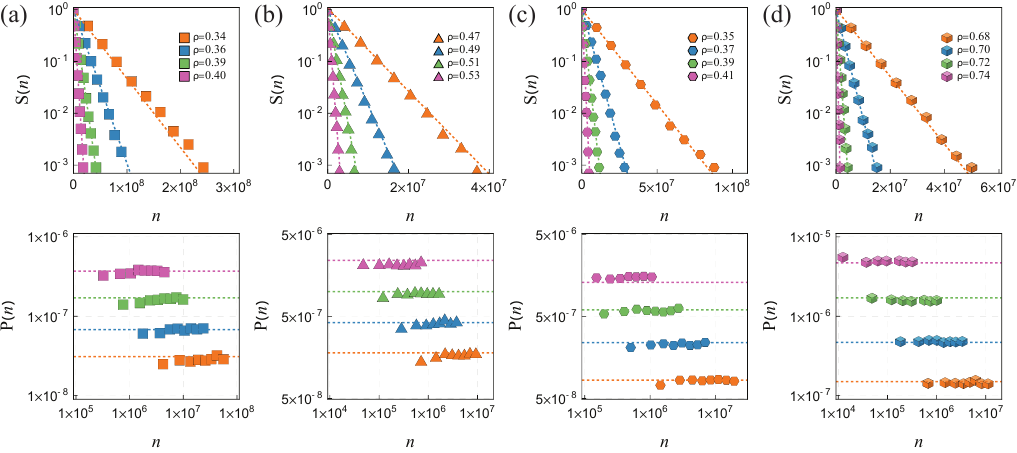}
\centering
\caption{\textbf{Exponentially decaying survival.} Top: measurements of $S(n)$ for various lattices. Bottom: Direct measurements of $P$ (markers) compared to the $P$ estimated via the exponential decay. Data is shown for (a) Square, (b) Triangular, (c) Hexagonal and (d) SC lattices.}\label{fig: trapping measurements appendix}
\end{figure}

 We estimated $S(n)$ for additional densities. We simulated $10^4$ walks for each density on each lattice. Results are shown in Fig. \ref{fig: trapping measurements appendix} top half. The exponential decay is clearly demonstrated. 

To estimate $P(n)$ directly, and check if it is constant or depends on $n$, we used estimates of $S(n)$ that were obtained from simulations. We divide those measurements into $m=10$ consecutive bins, such that the number of tracers that were trapped in each bin is identical and given by $M\approx 10^4/m=10^3$, yielding stable statistics. We used each such bin to estimate the mean trapping probability at every step during the time interval associated with the relevant bin using

\begin{equation} \label{eq:P calculation}
P\left(\frac{n+n'}{2}\right)\approx1-\left(\frac{S(n')}{S(n)} \right)^{\frac{1}{n'-n}}.
\end{equation}
\noindent where $n'$ is the number of steps at the start of the bin and $n$ is the number of steps at the end of the bin.

Next, we compare the estimated rate of the exponential decay of the survival with the directly measured $P(n)$, as shown in the bottom half of Fig. \ref{fig: trapping measurements appendix}. We find good agreement across obstacle densities and on several different lattices.

\section{Derivation of Eq. (3)}\label{appendix: trapping derivation}
\begin{figure}[h!]
\includegraphics[width=0.96\textwidth]{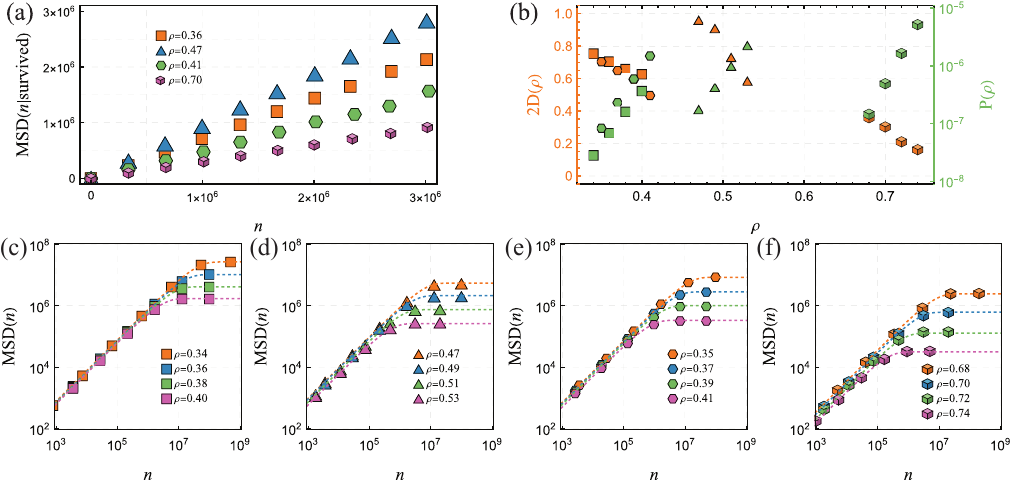}
\centering
\caption{\textbf{Effective $D(\rho)$ and $P(\rho)$ determine $\mathrm{MSD}(n)$.} (a) Simulation results of $\mathrm{MSD}(n|\text{survived})$ for walks that are not confined by step $n$. (b) Measurements of $D(\rho)$ (orange) and $P(\rho)$ (green) for various lattices and densities. The resulting values were used to predict the $\mathrm{MSD}(n)$ on the (c) Square (d) Triangular (e) Hexagonal and (f) Simple Cubic lattices.}\label{fig: d,q,MSD appendix}
\end{figure}

To begin, we divide the $\mathrm{MSD}(n)$ into two contributions,
\begin{equation} \label{eq: MSD _t_ total prob}
\mathrm{MSD}(n)=\mathrm{MSD}_{\text{survivors}}(n) + \mathrm{MSD}_{\text{trapped}}(n).
\end{equation}
The contribution of the survivors is 
\begin{equation} \label{eq: MSD survivors con}
\mathrm{MSD}_{\text{survivors}}(n)=\mathrm{MSD}(n|\text{survived})S(n).
\end{equation}

\noindent In Fig. \ref{fig: d,q,MSD appendix}, we show that when conditioned on survivors the MSD increases linearly in $n$, with a $\rho$ dependent diffusion constant: $D(\rho)$. This gives us the first term. The second term, the survival, is exponentially decaying as we show in Fig. \ref{fig: trapping measurements appendix}. Therefore we can write

\begin{equation} \label{eq: MSD survivors con explicit}
\mathrm{MSD}(n|\text{survived})=2nD(\rho); \quad\mathrm{MSD}_{\text{survivors}}(n)=2nD(\rho)e^{-nP(\rho)}.
\end{equation}

For the contribution of those already trapped by step $n$ we only need to integrate

\begin{equation}
\begin{split} \label{eq: MSD trapped con}
\mathrm{MSD}_{\text{trapped}}(n)=\int^{n}_0 \mathrm{MSD}(n'|\text{survived})\left(-\frac{d(S(n'))}{dn'}\right)dn'=\frac{2D(\rho)}{P(\rho)}\left(1 -(1+nP(\rho))e^{-nP(\rho)} \right).
\end{split}
\end{equation}

The same result can be derived more rigorously by computing the expectation as a sum

\begin{align} 
\text{MSD}_{\text{trapped}}(n) &= \sum_{n'=0}^{n} 2n' D(\rho) P(\rho) (1 - P(\rho))^{n'} \label{eq:MSD_trapped_con} \\
&= \frac{2D(\rho)}{P(\rho)} \left( 1 - (1 + nP(\rho))(1 - P(\rho))^n \right) \\
&\approx \frac{2D(\rho)}{P(\rho)} \left( 1 - (1 + nP(\rho)) e^{-nP(\rho)} \right)
\end{align}

We can plug both contribution into Eq. (\ref{eq: MSD _t_ total prob}) to get

\begin{equation}
\begin{split}
\label{eq: MSD _t_ derivation, explicit contributions}
\mathrm{MSD}(n)= 2D(\rho)n e^{-nP(\rho)}+2D(\rho)\left(\frac{1}{P(\rho)} -\left(\frac{1}{P(\rho)}+n \right)e^{-nP(\rho)} \right) =2D(\rho)(1-e^{-nP(\rho)})/P(\rho).
\end{split}
\end{equation}
To verify this result, we extend Fig. \ref{fig: MSD _t_} to additional densities. We begin by estimating $D(\rho)$ and $P(\rho)$, using simulations. In these simulations, we used $10^4$ walkers for each density on each lattice. To measure $D(\rho)$ we condition the simulation on the tracer not being trapped. To estimate $P(\rho)$ we find that it is enough to fit the exponential decay in $S(n)$ at short times, which emphasizes that only short-time information is required and makes the procedure computationally cheap. The $D(\rho)$ and $P(\rho)$ we obtained are shown in Fig. \ref{fig: d,q,MSD appendix}. Next, we used these constants to extend Fig. \ref{fig: MSD _t_} to additional densities, as shown in Fig. \ref{fig: d,q,MSD appendix}. Excellent agreement is found between simulation (symbols) and the predictions coming from Eq. (3), which are given by the dashed lines.

\section{Comparison with Singh et al.}\label{appendix: comparison}

\begin{figure}[h]
\includegraphics[width=0.96\textwidth]{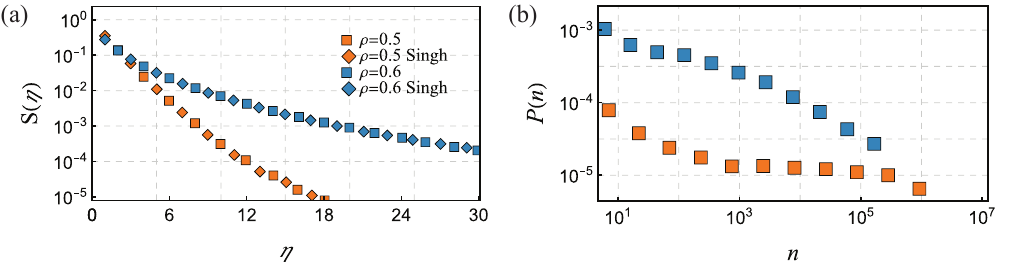}
\centering
\caption{\textbf{Comparison with Singh et al.} (a) Survival probability for two versions of the \textit{Sokoban} random walk on the 2D Square lattice with high obstacle densities: (i) without allowing failed moves as done in this paper (squares); (ii) with failed moves as done in \cite{Singh2026,singh2026-2} (tilted squares). When plotted against the normalized time $\eta=n/\langle n \rangle$, both dynamics collapse onto the same survival curve. (b) Direct measurements of the trapping probability $P(n)$ for the same densities. No plateau is observed; instead, $P(n)$ decreases monotonically with $n$.}
\label{fig: Eli appendix, high densities}
\end{figure}

Very recently, Singh et al. studied a variant of the \textit{Sokoban} random walk in which failed moves count as time steps \cite{Singh2026,singh2026-2}. In their model, the tracer chooses a random direction at each step, regardless of whether motion in that direction is possible. As a result, attempts to move into blocked configurations contribute as waiting times to the dynamics. The work of Singh et al. focused on high obstacle densities ($\rho>0.45$), i.e., above the regular percolation threshold for the 2D Square lattice. 

To compare directly with the results of Singh et al., we simulated both versions of the dynamics at the same high obstacle densities using $10^6$ walk realizations, allowing us to probe the survival probability over five orders of magnitude. Figure \ref{fig: Eli appendix, high densities}(a) shows that tallying waiting times due to failed move attempts does not qualitatively affect the kinetics once time is rescaled by the mean trapping time $\eta=n/\langle n \rangle$. At these high densities, we indeed observe strong deviations from exponential kinetics, in agreement with the observations of Singh et al. This is further confirmed in Fig. \ref{fig: Eli appendix, high densities}(b), where the instantaneous trapping probability $P(n)$ is shown to decrease monotonically with $n$ and no plateau is observed.

\section{Large scale simulations at low obstacle densities}\label{appendix: low-density}

\begin{figure}[h]
\includegraphics[width=0.96\textwidth]{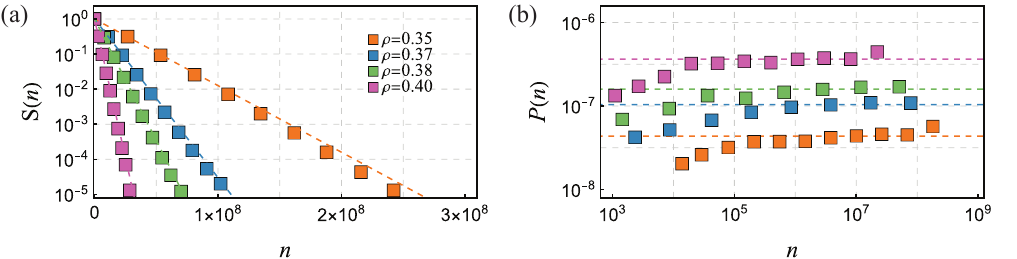}
\centering
\caption{\textbf{Effective exponential trapping kinetics at low obstacle densities.} (a) Measurements of $S(n)$ for densities below the regular (no-pushing) percolation threshold (markers) together with exponential fits (dashed lines). (b) Direct measurements of $P(n)$ (markers) compared with the constant decay rate extracted from the fits in panel (a). After a short transient, an approximately constant trapping probability is observed.} 
\label{fig: Eli appendix, low densities}
\end{figure}

In the low-density regime studied throughout this work, i.e., near and below the no-pushing percolation threshold, the trapping kinetics differ qualitatively from the high-density regime discussed in Appendix \ref{appendix: comparison}. In this appendix, we present large scale simulations of the \textit{Sokoban} for these lower densities ($10^6$ walk realizations vs. $10^4$). Figure \ref{fig: Eli appendix, low densities}(a) shows that, after a short transient, the survival probability is well described by exponential decay over several orders of magnitude in $S(n)$. To verify this directly, we measured the instantaneous trapping probability $P(n)$, shown in Fig. \ref{fig: Eli appendix, low densities}(b). At short times, $P(n)$ increases with $n$, reflecting the gradual formation of trapping configurations. However, this growth slows significantly at later times, where an extended plateau emerges. In this regime, the trapping probability is approximately constant, yielding effective exponential survival kinetics.

The distinction between the low- and high-density regimes is therefore not merely quantitative. In the high-density regime (Fig. \ref{fig: Eli appendix, high densities}(b)), the trapping probability decreases continuously with time and no plateau is observed. Conversely, in the low-density regime, $P(n)$ approaches a nearly constant value after a transient period where it monotonically increases. These observations suggest that the two regimes are governed by different dynamical mechanisms.

\begin{figure}[t]
\includegraphics[width=0.96\textwidth]{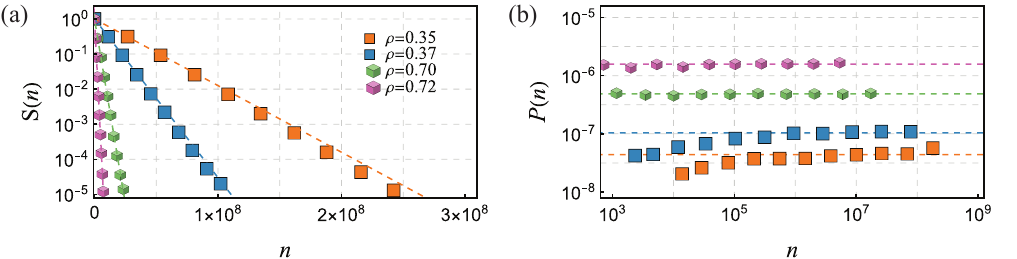}
\centering
\caption{\textbf{Geometry dependence of the trapping kinetics.} (a) Measurements of $S(n)$ on the Square and Simple Cubic lattices (markers) together with exponential fits (dashed lines). (b) Direct measurements of $P(n)$ (markers) and the corresponding constant decay rates extracted from panel (a). While the Square lattice displays a slow monotonic increase in $P(n)$, this trend is absent on the Simple Cubic lattice.} 
\label{fig: Eli appendix, 3D 2D}
\end{figure}

At longer times, we observe weak but measurable deviations from purely exponential kinetics in the 2D Square lattice. Fitting the survival probability to the form $S(n)=ae^{-Pn^c}$ yields $c\approx1.02$, indicating a slow increase in the trapping probability with time.

To further probe the time dependence of the trapping probability, we also performed simulations with $10^6$ walkers on the 3D Simple Cubic lattice. Results are shown in Fig. \ref{fig: Eli appendix, 3D 2D}. While the Square lattice exhibits a slow monotonic increase in $P(n)$, no comparable trend is observed on the Simple Cubic lattice. Thus, effective exponential kinetics provide an excellent description in 2D, and an almost perfect description in 3D, over the numerically accessible regime.

\end{document}